\documentclass[prl,twocolumn,showpacs,preprintnumbers,amsmath,amssymb,superscriptaddress]{revtex4}

\usepackage{graphicx}
\usepackage{dcolumn}
\usepackage{bm}

\usepackage{amsmath}	
\begin{document}
\draft
\newcommand{\lw}[1]{\smash{\lower2.ex\hbox{#1}}}

\title{Hole Localization in One-Dimensional Doped Anderson--Hubbard Model}

\author{M.~Okumura}
\email{okumura.masahiko@jaea.go.jp}
\affiliation{CCSE, Japan Atomic Energy Agency, 6--9--3 Higashi-Ueno,
Taito-ku Tokyo 110--0015, Japan}
\affiliation{CREST (JST), 4--1--8 Honcho, Kawaguchi, Saitama 332--0012,
Japan}
\author{S.~Yamada} 
\email{yamada.susumu@jaea.go.jp}
\affiliation{CCSE, Japan Atomic Energy Agency, 6--9--3 Higashi-Ueno,
Taito-ku Tokyo 110--0015, Japan}
\affiliation{CREST (JST), 4--1--8 Honcho, Kawaguchi, Saitama 332--0012,
Japan}
\author{N.~Taniguchi}
\email{taniguch@sakura.cc.tsukuba.ac.jp}
\affiliation{Institute of Physics, University of Tsukuba, Tennodai
Tsukuba 305--8571, Japan}
\author{M.~Machida}
\email{machida.masahiko@jaea.go.jp}
\affiliation{CCSE, Japan Atomic Energy Agency, 6--9--3 Higashi-Ueno,
Taito-ku Tokyo 110--0015, Japan}
\affiliation{CREST (JST), 4--1--8 Honcho, Kawaguchi, Saitama 332--0012,
Japan}

\date{\today}

\begin{abstract} 
  
 We study an interplay of disorder and correlation in the
 one-dimensional hole-doped Hubbard-model with disorder 
 (Anderson-Hubbard model) by using the density-matrix renormalization  
 group method. Concentrating on the doped-hole density profile, we find 
 in a large $U/t$ regime that the clean system exhibits a simple
 fluidlike behavior whereas finite disorders create locally Mott
 regions which expand their area with increasing the disorder strength 
 contrary to the conventional sense. We propose that such an anomalous
 Mott phase formation assisted by disorder is easily observable in
 atomic Fermi gases by setup of the box-shape trap. 

\end{abstract}
\pacs{71.10.Fd, 71.10.Pm, 71.23.-k, 03.75.Ss}

\maketitle

Recently, atomic Fermi gas loaded on an optical lattice (FGOL) has
attracted a lot of attention, since FGOL is expected to be an excellent  
testbed to resolve controversial issues in condensed matter physics
\cite{OLtheory}. One of the advantages of FGOL is a tunability of the
interaction between two atoms associated with the Feshbach resonance
\cite{Feshbach}, which opens up a pathway to systematically study not
only BCS-BEC crossover but also strongly correlated behaviors. Another
advantage is the flexibility in making playgrounds such as the
periodical lattice, which provides various stages including disorder
effects for many-body interacting systems \cite{OLtheory}. 

Among a huge number of proposals on FGOL, one of the unique challenges
is a study of interplay between randomness and strong correlation
\cite{OLtheory}. This is one of the most difficult but important problems 
in real solids because high-$T_{\rm c}$ superconductor is a typical
reality. In high-$T_{\rm c}$ superconductors, their common mother phase
is the Mott insulator showing antiferromagnetism \cite{dopedMott}. The
carrier is doped by chemical substitution, which inevitably brings a
random potential. However, the disorder effects in strongly correlated
systems have been too complicated issues to study theoretically and
experimentally in condensed matters. Thus, its interplay has remained as
an unsolved issue. On the other hand, FGOL is a very good experimental
reality in systematically examining such a complex issue due to the wide
tunability and flexibility. 

In this paper, we study the doped Mott insulator with disorder in a form 
of the Anderson-Hubbard model [see Eq.~(\ref{Hamiltonian}) below], and
predict experimental results on FGOL by means of the density-matrix
renormalization group (DMRG) method \cite{White,DMRGreview}. 
Consequently, we find that the disorder does not destroy the Mott
insulator but help the growth of the Mott phase domains contrary to our 
naive expectation. Such a nontrivial feature is kept and amplified
until the disorder amplitude fully exceeds over the repulsive
interaction strength. 

So far, the harmonic trapped FGOL has been considered in
theoretical studies. For instance, Gao {\it et al.} have reported their
DMRG calculation results for the 1D Anderson--Hubbard model
\cite{Anderson,Hubbard} with the harmonic trap potential \cite{Gao}. 
However, the harmonic trap induces spatially inhomogeneous filling which 
is different from usual situations in solid state matters. On the other
hand, the box trap with disorder provides almost equivalent stages. 
Thus, our target reality is one-dimensional (1D) FGOL confined in a
box-shaped trap with lattice including randomness. In its experimental
setup, see Ref.~\cite{Raizen} for the box-shape trap and
Ref.~\cite{OLtheory} for the random potential. 

The Hamiltonian of the 1D Anderson--Hubbard model is given by
\begin{equation}
 H_{\rm AH} = -t \sum_{\langle i,j \rangle, \sigma} c^\dag_{i\sigma}
 c_{j\sigma} + \sum_{i, \sigma} \epsilon_i n_{i,\sigma} + \sum_{i}
 U n_{i\uparrow} n_{i\downarrow}\, , \label{Hamiltonian}
\end{equation}
where $\langle i ,j \rangle$ refers to the nearest neighbors $i$ and
$j=i\pm1$, $t$ is the hopping parameter between the nearest neighbor
lattice sites, $U$ is the on-site repulsion, $c_{i\sigma}$
($c_{i\sigma}^\dag$) is the annihilation (creation) operator with spin
index $\sigma=\uparrow, \downarrow$, $n_{i,\sigma} (\equiv
c^\dag_{i\sigma} c_{\sigma})$ is the site density operator, and the
random on-site potential $\epsilon_i$ is chosen by a box probability
distribution ${\cal P} (\epsilon_i) = \theta ( W/2 - |\epsilon_i|)/W$,
where $\theta (x)$ is the step function and the parameter $W$ controls
the disorder strength. In all DMRG calculations, we employ the open
boundary condition for the box-shape trap, and focus on the site density
of fermions as $n_i = n_{i,\uparrow} + n_{i,\downarrow}$.

The 1D Anderson--Hubbard model has been intensively investigated in the
context of the transition between the Anderson and Mott insulators
\cite{1DAH,ALreview,1DAHring}.  To our knowledge, however, this model
has not been fully studied in a range of slight to under doping. 
Although the model may look simple, the interplay among the disorder,
the interaction, and the doped holes requires more accurate and more 
systematic studies.  This paper provides the first systematic results of 
doped-hole profiles under a full accuracy.   

In order to calculate the ground state of the Hamiltonian
(\ref{Hamiltonian}), we use the DMRG method \cite{White,DMRGreview}. 
The validity of our DMRG results was verified by results of the exact
diagonalization in a case of $(8\uparrow, 8\downarrow)$ with $18$ sites 
($L=18$) at $U/t=W/t=30$. Both the results give a good agreement. The
size dependence in terms of doped-hole profiles is checked for $L=50$,
$100$, $150$, and $200$, which reveal that $L=50$ is enough to
characterize hole profiles. In the following, we present results of 1D
system with the length $L=50$ in three fillings, $\bar{n}=\sum_{i=1}^{L}
n_i/L =0.96$, $0.88$, and $0.52$. In the use of DMRG, the number of
states kept $m$ is set maximum 256 for several cases, and $m=100$ is
confirmed to be enough to obtain convergent results because the largest
deviation of the local density between them is below $10^{-4}$. 

\begin{figure}
\includegraphics[scale=0.43]{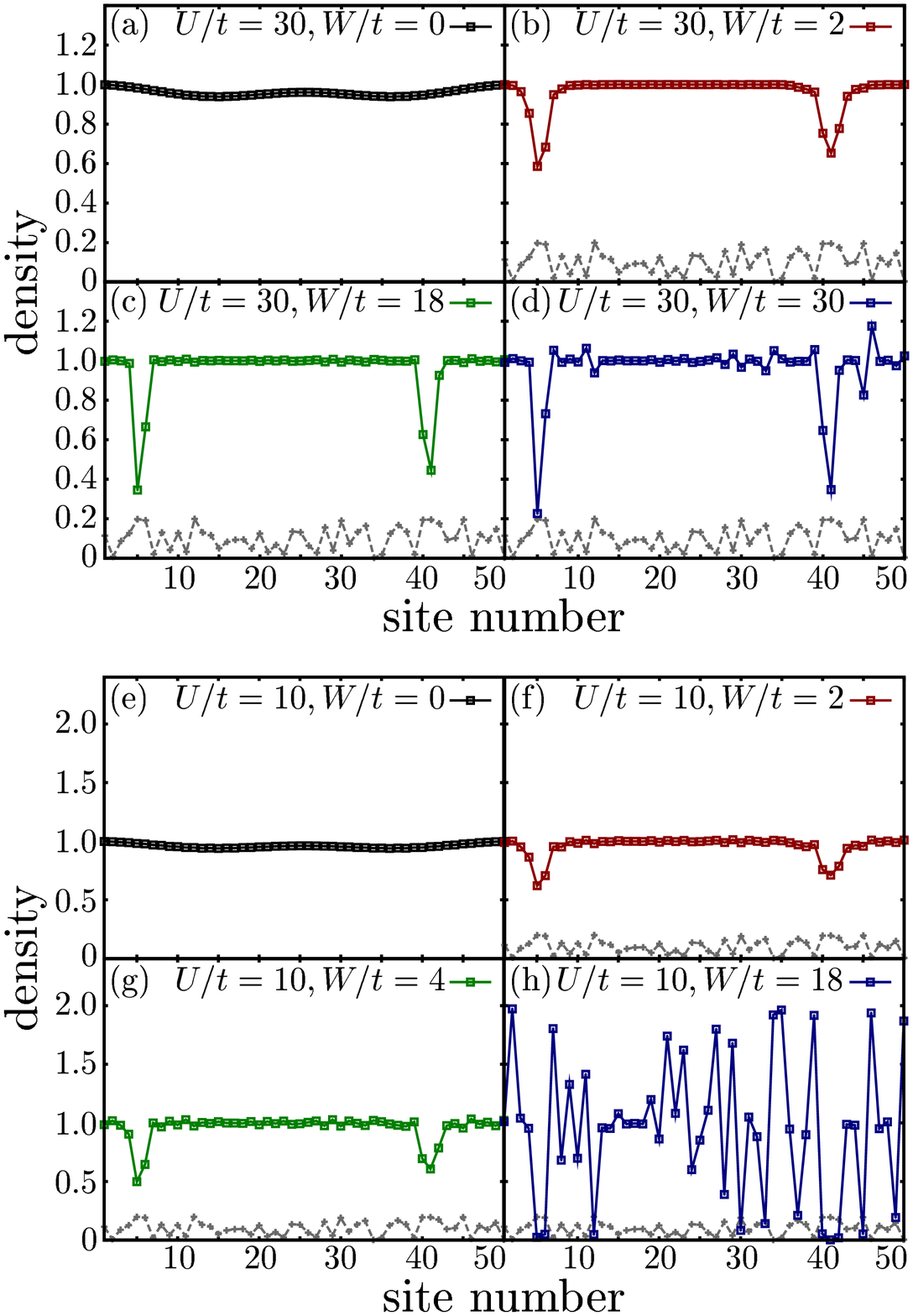}
\caption{\label{fig1} The randomness $W$ dependent density profiles
 $n(i)$ for 2-holes doped case ($\bar{n}=0.96$) at two fixed interaction
 strengths, (a)--(d) for $U/t=30$, and (e)--(h) for $U/t=10$, under a
 random potential depicted in the bottom of each figure in arbitrary
 unit (grey dashed line).} 
\end{figure}

First, let us show DMRG results of density profiles when 2 holes
($\uparrow, \downarrow$) are doped and the filling $\bar{n}=0.96$.
Figures \ref{fig1}(a)--\ref{fig1}(d) display typical profiles obtained
by varying $W$ in a fixed $U/t=30$ under a certain random configuration
as depicted at the bottom. In the clean limit ($W=0$), one finds a
fluidlike density profile as Fig.~\ref{fig1}(a), which belongs to the
Tomonaga--Luttinger liquid with open boundary conditions. With switching 
on $W$, flat density regions whose site density equals to a unit emerge
with two clear dips as Fig.~\ref{fig1}(b). The locations of the dips do
not shift by choice of the boundary condition, i.e., open or periodic
boundary condition. We also confirm the fact by using the exact
diagonalization in $L=18$ sites. Here, we name the flat density region
and the dip ``Mott plateau'' and ``hole valley'', respectively. We note
that the number of observed hole valleys equals to that of doped
holes. This implies that holes tend not to collect but to localize
separately. Furthermore, one finds from Figs.~\ref{fig1}(b) and
\ref{fig1}(c) that the edges of the Mott plateaus become sharper and the
hole valleys become deeper as the randomness strength $W$ increases from
$W/t=2$ to $18$ [see also Figs.~\ref{fig3} (b) and \ref{fig3}(c) for
another doping case]. This clearly indicates that the randomness
\emph{assists} the formation of Mott plateaus rather than breaking the
structure. This is quite nontrivial because disorder is usually believed
to break flat homogeneous distribution. Since the clean system exhibits
Tomonaga--Luttinger liquid except for the half-filling, these results
clearly illustrate that the randomness is essential for the local
development of the Mott state, where the strong interaction and the
randomness cooperatively form the Mott plateaus with localizing the
doped holes. This is suggestive for the fact that the insulator phase in
high-$T_{\rm c}$ superconductors survives tiny doping. 

When the randomness $W$ increases and approaches $W\sim U$, the Mott
plateaus are disturbed partly by disorder [see Fig.~\ref{fig1}(d)]. 
By further increase of $W$ above $U$, the Mott plateau with localized
holes breaks down. Figures \ref{fig1}(e)--\ref{fig1}(h) show typical
density profiles in a weaker interaction $U/t=10$. Similar to
Figs.~\ref{fig1}(b) and \ref{fig1}(c), the structure of the Mott plateau
with localized holes is seen for $W<U$, as Figs.~\ref{fig1}(f) and
\ref{fig1}(g). The structure is, however, completely destroyed by the
strong random potential $W > U$ as seen Fig.~\ref{fig1}(h). Under the
strong randomness, the local density takes values close to 2 (the double
occupation) at several sites, and the Mott plateaus are too small to be
identified as a region.

The $W$ dependent density profiles seen in Fig.~\ref{fig1} suggest that
there is an optimum $W$ $(U)$ for a fixed $U$ $(W)$ in making the Mott
plateaus as wide as possible and the hole valleys as sharp as possible. 
To evaluate an optimum set of parameters $(U,W)$, we introduce a
function $M(U,W)$ in the two variable space to characterize the extent
of the Mott plateau in the total density profile, which is defined by a 
summation of ``closeness'' of the local density to unit expressed as
\begin{equation}
 M(U,W) = \left\langle \sum_{i=1}^{L} \exp\left[
 \frac{-(n_{i}(\epsilon,U,W)-1)^2}{2\Delta^2} \right] / (\bar{n}L)
 \right\rangle_\epsilon , \label{MUW} 
\end{equation}
where $n_{i}(\epsilon,U,W) \equiv n_{i,\uparrow} + n_{i,\downarrow}$
means the local site density under a realization of the random potential
at a certain set of $U$, $W$ and a local potential $\epsilon$, and
$\langle \cdot \rangle_\epsilon$ is the algebraic average for multiple
random realizations. $\Delta$ characterizes the peak width of the
function, for which we fix $\Delta = 0.05$. We note that the factor  
$1/(\bar{n}L)$ is the normalization constant which is adjusted to make
$M(U,W)\simeq1$ when the extent of the Mott plateaus is the maximum,
e.g., $M(U,W) \simeq 1$ when $n_i=0$ at two lattice points and $n_i=1$
at other 48 lattice points if two holes are doped in $L=50$. 

\begin{figure}
\includegraphics[scale=0.42]{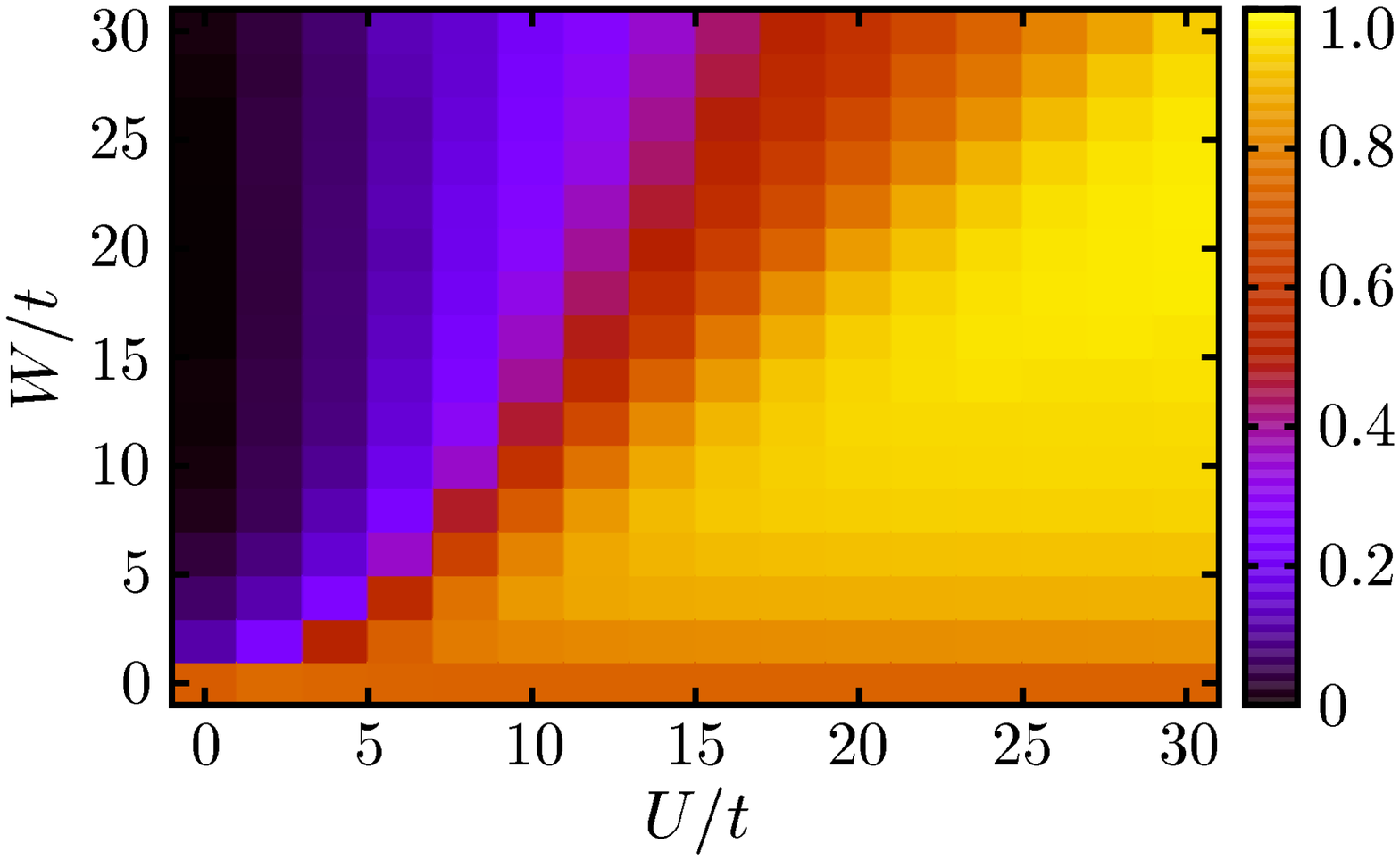}
\caption{\label{fig2} A contour plot of $M(U,W)$ in 2 holes doped case
  ($\bar{n}=0.96$) as a function of $U/t$ and $V/t$.  Arithmetic
  average over ten realizations of random potentials is taken.  The
  step values are 2 for both $U/t$ and $W/t$ axes.}
\end{figure}

\begin{figure}
\includegraphics[scale=0.43]{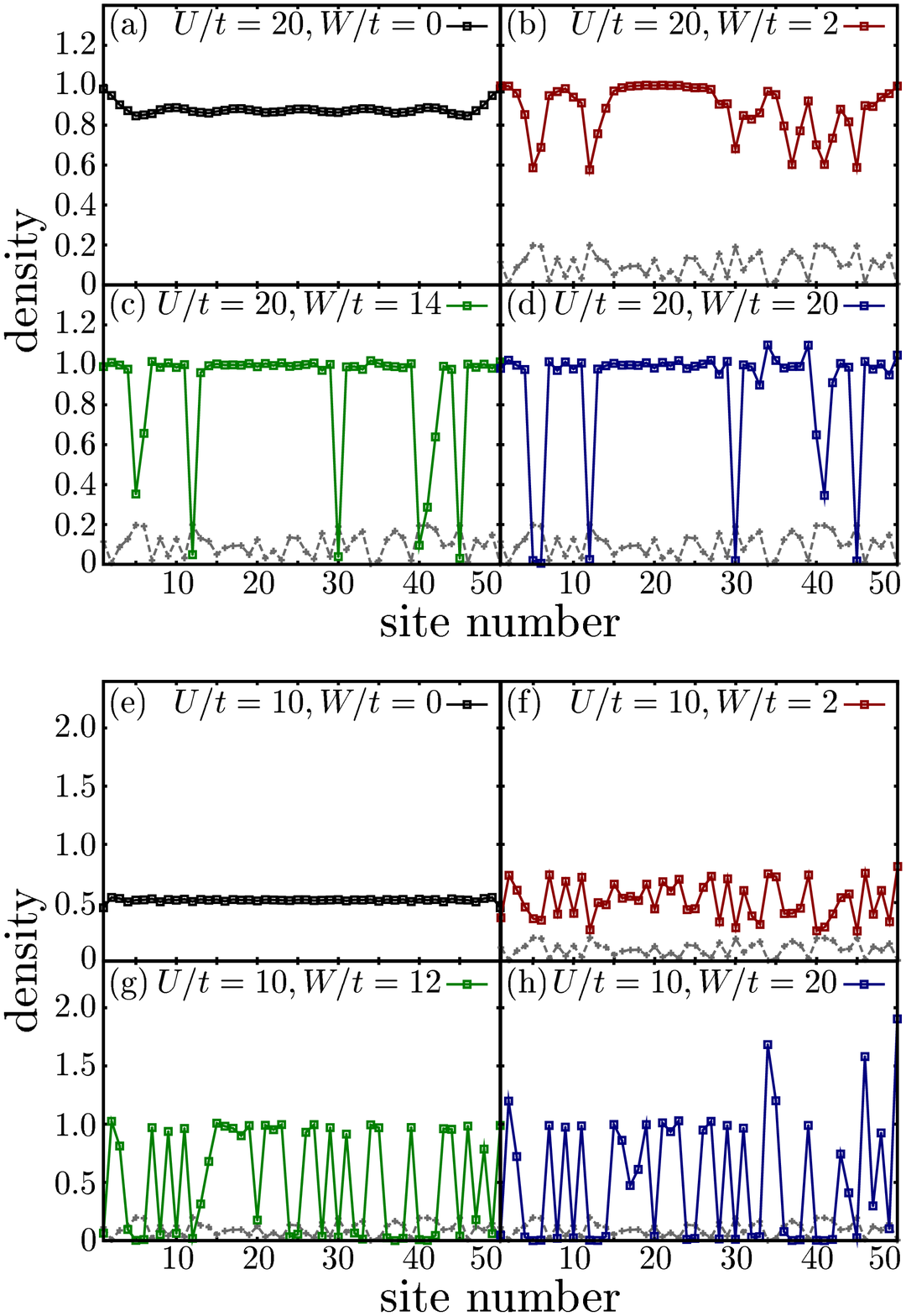}
\caption{\label{fig3} The randomness $W$ dependent density profiles
 $n(i)$ for (a-d) $6$ ($\bar{n}=0.88$) and (e-h) $24$ ($\bar{n}=0.52$)
 holes doped cases under a random potential shown in the bottom of each
 figure in arbitrary unit (grey dashed line). 
}
\end{figure}

Figure~\ref{fig2} shows a contour plot of $M(U,W)$ obtained by the
arithmetical average over ten realizations of random potentials for 
$\bar{n}=0.96$ (2 holes doped) case.  Along a fixed $W/t$ line, one
finds that the value of $M(U,W)$ increases monotonically with increasing 
$U/t$. This is consistent with what one sees in Fig.~\ref{fig1}, i.e.,
the width of the Mott plateaus in $U/t=30$ case (Fig.~\ref{fig1} (c))
is wider than that in $U/t=10$ case (Fig.~\ref{fig1} (h)) on the same 
randomness strength $W/t=18$.  On the other hand, along a fixed $U$
line, the value of $M(U,W)$ initially increases with increasing $W$,
reaches the maximum values close to $W +2t \sim U/2$ line, and then
decreases after crossing the maximum value. Thus, one finds in the
slightly-doped case that holes are confined as compactly as possible
when $W$ is about a half of $U$ so that the width of the hole valley is
almost a unit lattice constant. It indicates that doped holes almost
lose their quantum delocalizing character.
  
Here, we turn our attention to the cases in which more holes are doped.
Figure \ref{fig3} (a)--(d) shows the total density distributions of
fermions in $\bar{n}=0.88$ (6 holes) case with $U/t=20$, where we can
confirm almost the same behaviors as $\bar{n}=0.96$ (2 holes)
case. Namely, the introduction of disorder results in the formation of
both the Mott plateaus and hole valleys, and further increase of $W$
makes the Mott plateaus wider and hole valleys deeper (compare
Fig.~\ref{fig3} (b) ($W/t=2$) with (c) ($W/t=14$)). One also finds that
the number of valleys is exactly the same as that of doped holes at
$W/t=2$ case while the number becomes smaller than that of doped holes
at $W/t=14$ and $W/t=20$. The tendency to create the Mott plateau still
remains even for far away from the half-filling.  Figure~\ref{fig3}
(e)--(h) is the density profile for $\bar{n}=0.52$ (24 holes, i.e.,
close to the quarter filling) at $U/t=10$.  Although the number of
valleys is no longer the same as that of doped holes in all cases
($W/t=2$, $12$, and $20$) and the number of fermions is not enough to
simply form the Mott plateau at this filling, the maximum density is
nearly a unit (see Fig.~\ref{fig3} (g) and (h)). This suggests even in
such a high doping level that both the interaction and randomness
cooperatively work.  

\begin{figure}
\hspace*{-2mm}
\includegraphics[scale=0.42]{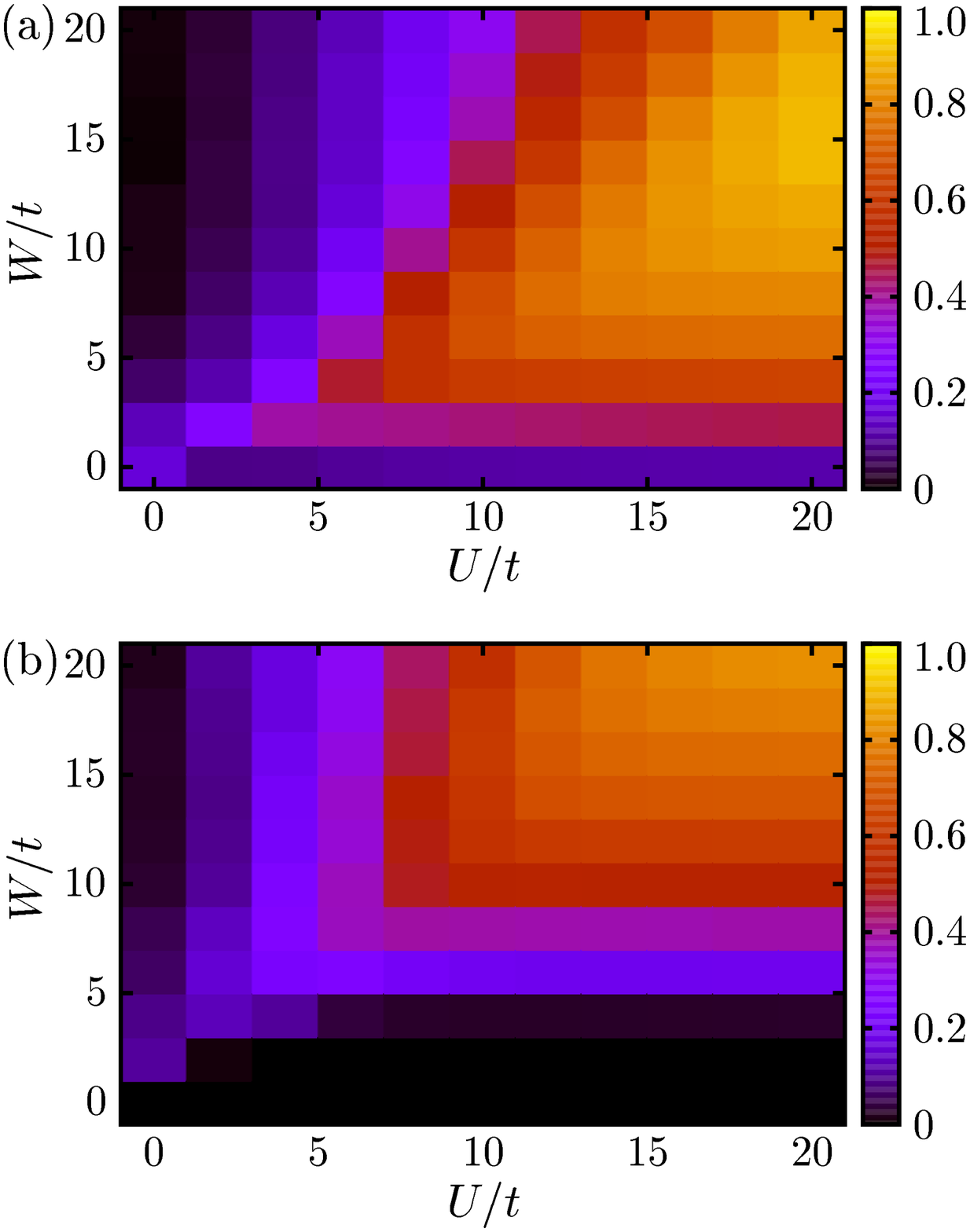}
\caption{\label{fig4} Contour plots of the value of $M(U,W)$ in (a)
 $\bar{n}=0.88$ (6 holes) and (b) $\bar{n}=0.52$ (24 holes) cases with
 the arithmetic average of ten realizations of random potentials. The
 step value for both $U/t$ and $W/t$ is 2.
 } 
\end{figure}

Figure \ref{fig4} (a) and (b) show contour plots of $M(U,W)$ for
$\bar{n}=0.88$ (6 holes) and $\bar{n}=0.52$ (24 holes) cases,
respectively. These essentially show the same tendency as $\bar{n}=0.96$ 
(2 holes) case.  The value of $M(U,W)$ increases with increasing $U/t$
at a fixed $W/t$, and the optimum value of $M(U,W)$ exists for a fixed
$U/t$. On the other hand, one finds in more details that the maximum of
$M(U,W)$ shifts to $W +2 t > U/2$ side with increasing the number of
doped holes. This is because, in high hole-density cases, the random
potential magnitude required to push the local density up to the unit
becomes larger. In addition, high hole-density causes the reduction of
the maximum value of $M(U,W)$, because it suppresses the correlation
effect and works unfavorably in forming the Mott plateaus. 

\begin{figure}
\includegraphics[scale=0.45]{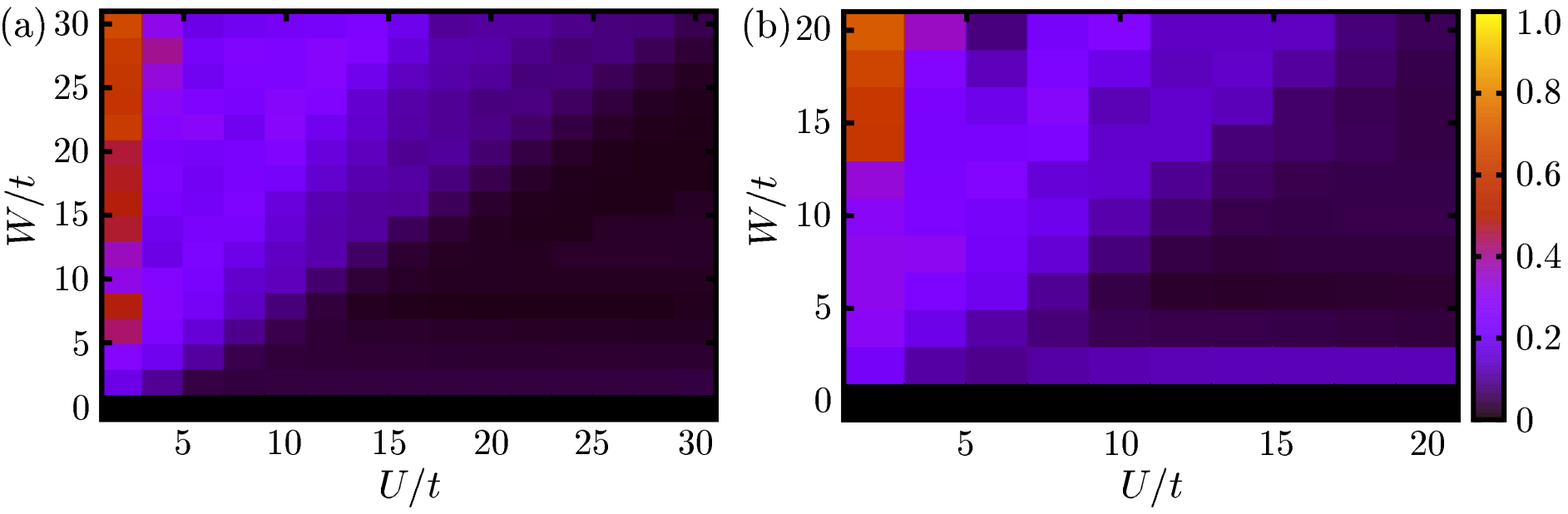}
\caption{\label{fig5} Contour plots of $D(U,W)$ in (a) 2 holes doped
 case ($\bar{n}=0.96$) and (b) 6 holes doped case ($\bar{n}=0.88$) as a
 function of $U/t$ and $V/t$.  Arithmetic average over ten realizations
 of random potentials is taken. The step values are 2 for both $U/t$ and
 $W/t$ axes.}
\end{figure}

We also examine fluctuations of $M(U,W)$ on different randomness by
evaluating the standard deviation per average $D(U,W) = \sqrt{ \langle
(M_\epsilon (U,W) / M (U,W) - 1)^2 \rangle_\epsilon}$, where $M_\epsilon
(U,W) = \sum_{i=1}^{L} \exp \left( - (n_i (\epsilon,U,W) - 1)^2 /
2\Delta^2 \right) / (\bar{n}L)$. As expected, one finds in
Fig.~\ref{fig5} (a) (2 holes doped case) that $D(U,W)$ becomes
relatively small when $M(U,W)$ shows large values, and vice versa. This
clearly demonstrates that the formation of Mott plateaus does not depend
on the realization of random potential and $M(U,W)$ is a good measure to
know it. We also obtain the almost same behavior of $D(U,W)$ in the 24
holes case as shown in Fig.~\ref{fig5} (b).

In conclusion, we calculated the density profiles of fermions in the 1D
Anderson--Hubbard model by varying interaction strengths, random
potential amplitudes, and fillings below the half-filling by means of
DMRG.  We found a clear signature that the presence of disorder assists
the local formation of the Mott state in the weak disorder region
whereas the Mott phase is destroyed by strong disorder. As a
characterization of the width of the Mott phase, we calculated the
function $M(U,W)$ from the density profiles, and found that the increase
of the doping rate shifts the maximum of $M(U,W)$ from $W + 2t = U/2$
line to $W + 2 t>U/2$ side. These non-trivial behaviors of doped holes
like the present DMRG works can be systematically examined by FGOL with 
the box trap. The experimental confirmation will give a crucial
contribution to studies for doped Mott insulators.

The authors wish to thank H.~Aoki, T.~Deguchi, K.~Iida, T.~Koyama,
H.~Matusmoto, Y.~Ohashi, T.~Oka, S.~Tsuchiya, and Y.~Yanase for
illuminating discussion. The work was partially supported by
Grant-in-Aid for Scientific Research (Grant No.~18500033) and one on
Priority Area ``Physics of new quantum phases in superclean materials''
(Grant No.~18043022) from the Ministry of Education, Culture, Sports,
Science and Technology of Japan. One of authors (M.M.) is supported by
JSPS Core-to-Core Program-Strategic Research Networks, ``Nanoscience and
Engineering in Superconductivity (NES)''. 


\end{document}